\documentstyle[pictex,preprint,aps,epsfig]{revtex}
\draft
\preprint{SNUTP 99-001, KIAS-P99001}
\begin{document}
\title{\Large\bf A natural solution of $\mu$ from the
hidden sector}
\author{Jihn E. Kim\footnote{jekim@phyp.snu.ac.kr}}  
\address{Department of Physics and Center for Theoretical
Physics, Seoul National University,
Seoul 151-742, Korea, and\\
School of Physics, Korea Institute for Advanced Study, 
Cheongryangri-dong, Dongdaemun-ku, Seoul 130-012, 
Korea}
\maketitle

\begin{abstract}
The $\mu$ parameter is calculated in supergravity models
possessing the $U(1)_A\times U(1)_R$ symmetry. 
In one natural model without a free mass parameter
below the Planck scale, the symmetry breaking scale
is identified as the hidden sector squark condensation
scale.
\end{abstract}

\def\a{${\ddot {\rm a}}$}

\newpage

\section{Introduction}

The $\mu$ problem \cite{mu} is one of the mass hierarchy
problem reintroduced in supersymmetric models
toward a gauge hierarchy solution. In the literature,
many mechanisms to resolve this unexpected $\mu$ were proposed 
\cite{mu,gm,ckn,musol}. 
Among these solutions, the existence of an underlying symmetry
to forbid a large $\mu$ term is most compelling \cite{kn}.
The so-called Giudice-Masiero mechanism \cite{gm} must
also assume a symmetry, otherwise the absence
of $M_{Pl}H_1H_2$ in the superpotential is not
guaranteed. A $U(1)$ global symmetry
or $R$ symmetry is enough to forbid the $\mu$ term at tree
level in the superpotential \cite{chun}. 

In supergravity, the K${\ddot {\rm a}}$hler 
potential $K$ and the superpotential $W$ have the symmetry
\begin{equation}
G=K+\log |W|^2={\rm invariant}.
\end{equation}
Therefore, even if the $\mu$ term is 
forbidden in $W$ by a symmetry (say, by the $R$ symmetry 
with $R=0$ for the $H_1H_2$ operator), one can write it
in $K$ if the symmetry allows it. {}For example, 
one may consider
$[\epsilon H_1H_2/M_P^2+{\rm h.c.}+\log |W|^2]$ which is the same as
$\log |W^\prime|^2$ through the relation (1) where
$$
W^\prime=We^{\epsilon H_1H_2/M_P^2}
$$
where $M_P=M_{Pl}/\sqrt{8\pi}=2.44\times 10^{18}$ GeV.
Thus the supergravity introduces nonrenormalizable terms suppressed
by $M_P$ and the $\mu$ term is generated at the level $<W>/M_P^2
\sim m_{3/2}$.
Is this enough to state that supergravity has a natural scale for
$\mu$ even without a symmetry principle?

If we require that the theory dictates no superpotential at some
basis, then any interaction must come from the K{\a}hler potential,
\begin{equation}
G=K_0+{1\over M_P^3}\left(W_0+\bar W_0\right)+\log |1|^2.
\end{equation}
By the transformation (1)
$
G=K_0+\log|e^{W_0/M_P^3}|^2,
$
we obtain a superpotential including definite nonrenormalizable
terms,
\begin{equation}
W=\Lambda^3e^{W_0/M_P^3}
\end{equation}
where $\Lambda$ is a mass parameter. Requiring that $W$ contains
the known Yukawa interactions, $\Lambda$ must be of order $M_P$,
implying $<F>\simeq M_P^2$ and $m_{3/2}\sim M_P$.
Writing the K{\a}hler potential as $G$ of Eq. (2),
$G$ must contain O$(M_P)H_1H_2$+h.c., if no global
symmetry is imposed in $G$.
Here
the mass parameter must be of order $M_P$. Then $W$ contains the
$\mu$ term of order $M_P$. {}From this example, we note that the
TeV scale $\mu$ term is not a generic feature from K{\a}hler
potential in supergravity.

Therefore, there is a need to define $W$ more clearly.
In this paper, we {\it define the superpotential $W$ as the maximum
obtained from the transformation (1)}, i.e. when there is no
piece left in the K{\a}hler potential which can be transformable
to $W$.  This is the reason why we must include all possible 
nonrenormalizable terms in $W$. In this basis, we can effectively
apply the nonrenormalization theorem in $W$. In the
following, we respect the symmetry both in the superpotential 
and K{\a}hler potential. The K{\a}hler potential may be split into
$K_1+K_2$ where $K_1$ is nonholomorphic and $K_2$ is holomorphic
in the sense $K_2=W_2+{\rm h.c.}$. Then we take $K_1$ as the K{\a}hler
potential and $K_2$ must be included in the superpotential.
If $K_2$ were not respecting a global symmetry of the original
superpotential, then our final superpotential would not respect the
global symmetry. Therefore, the symmetry principle we impose on
the superpotential must apply also to the K{\a}hler potential.
Under this symmetry principle, various possibilities of generating
the electroweak scale $\mu$ were considered before \cite{kn}.

\section{The $U(1)_A\times U(1)_R$ Symmetry}

Two most popular global symmetries in supersymmetric models 
are the Peccei-Quinn symmetry $U(1)_A$ and $U(1)_R$ symmetry. 
Of course, the symmetries are  respected by
the K{\a}hler potential, and
there is no piece left in $K$ which can be transformed to $W$.
In this spirit, we must include all nonrenormalizable terms in $W$.  

We impose the symmetry $U(1)_A\times U(1)_R$. The relevant
fields for our purpose are listed in Table~1 with their quantum
numbers.

\vskip 0.5cm
\centerline{Table 1. The $A$ and $R$ quantum numbers of superfields.}
\vskip 0.5cm
\begin{center}
\begin{tabular}{|c|c|c|c|c|c|c|c|}
\hline
&\ \ $H_1$\ \ &\ \ $H_2$\ \ &\ \ $S_1$\ \ &\ \ $S_2$\ \ &\ \  
$S_3$\ \ &\ \ $S_4$\ \ &\ \ $Z$\ \ \\
\hline
\ \ $A$\ \  &--1&--1&1&--1&2&--2&0\\
\ \ $R$\ \  &0&0&1&--1&2&0&2\\
\hline
\end{tabular}
\end{center}
\vskip 0.5cm

\noindent The most general superpotential consistent
with the $U(1)_A\times U(1)_R$ symmetry, for $d\le 4$, is
\begin{equation}
W=f_0H_1H_2S_3+f_1Z(S_1S_2-F^2)+MS_3S_4+f_2S_1^2S_4+{f_3\over M_P}H_1
H_2S_1^2
\end{equation}
where $F$ is a mass parameter at $10^{12-13}$ GeV scale
and $M$ is of order $M_P$.
Thus, when $S_1$ obtains a vacuum expectation value
of order $F$, we obtain a reasonable $\mu$
\begin{equation}
\mu={f_3\over M_P}F^2.
\end{equation} 

In this scheme, there is no contribution to $\mu$ from the
K{\a}hler potential. To have a contribution to $\mu$ {\it a la}
Giudice and Masiero \cite{gm}, we must introduce
a singlet field $S_5$ with $A=-2, R=0$, so that
the K{\a}hler potential includes $S_5^*H_1H_2+{\rm h.c.}$
Then we obtain
\begin{equation}
\int d^2 \bar\theta\int d^2\theta {1\over M_P}S_5^*H_1H_2=\int d^2
\theta {F_{S_5^*}\over M_P}H_1H_2
\end{equation}
from which $\mu$ term is interpreted as $F_{S_5^*}/M_P$.
If the F-term of $S_5$ is nonvanishing and is of order $10^{11}$~GeV,
then we obtain a correct order of $\mu$. This necessarily assumes
a supersymmetry breaking mechanism, which is contrasted to $\mu$
arising from VEV of a scalar field given in Eq.~(5).

\section{A Natural Model for $\mu$}

In the above example, the intermediate scale parameter $F$ is
inserted by hand.  Also, if the Giudice-Masiero mechanism is to be
introduced, a specific form of the supersymmetry breaking at the 
intermediate scale must be assumed. Therefore, we must include the
intermediate scale physics. Moreover, it is natural that
if any gauge singlet is introduced, the mass parameter accompanying
the singlet is of order of the Planck scale.
Thus it is better if this scale $F$ of Eq.~(4) derives
from the h-sector confining force, rather than
putting it by hand in the superpotential. Along this line,
we introduce an h-sector
gauge group as $SU(N)_h$. In Ref.~\cite{ckn}
this idea has been proposed to generate a $\mu$ term from
the hidden sector squark (h-squark) condensation
through the nonrenormalizable term,
$$
{1\over M_P}H_1H_2\bar Q_1Q_2.
$$
But, in Ref.~\cite{ckn}, it was not given how $\bar Q_1Q_2$
develops a vacuum expectation value. 

We proceed to discuss to generate $<\bar Q_1Q_2>$ from the
intermediate scale physics.  
Let us consider the fields given in Table~2
with the $U(1)_A\times U(1)_R$ symmetry. We introduce
two chiral h-quarks $Q_2, Q_4$ and two chiral anti-h-quarks
$\bar Q_1,\bar Q_3$. {}For the h-sector $SU(N)_h$ gauge group, these
transfom as $N$ and $N^*$, respectively.

\vskip 0.5cm
\centerline{Table 2. The $A$ and $R$ quantum numbers with h-quarks $Q_i$.}
\vskip 0.5cm
\begin{center}
\begin{tabular}{|c|c|c|c|c|c|c|c|c|}
\hline
&\ \ $H_1$\ \ &\ \ $H_2$\ \ &\ \ $\bar Q_1$\ \ &\ \ $Q_2$\ \ &\ \  
$\bar Q_3$\ \ &\ \ $Q_4$\ \ &\ \ $S$\ \ &\ \ $S^\prime$\ \ \\
\hline
\ \ $A$\ \  &--1&--1&1&1&1&1&--2&2\\
\ \ $R$\ \  &0&0&1&1&--1&--1&0&2\\
\hline
\end{tabular}
\end{center}
\vskip 0.5cm
The $d\le 4$ superpotential consistent with the symmetry is
\begin{equation}
W=MSS^\prime+H_1H_2S^\prime+\bar Q_1Q_2S+{1\over M_P}H_1H_2\bar Q_1Q_2
\end{equation}
where we suppressed the couplings of order 1.
Due to the symmetry, there cannot appear $\bar Q_1Q_2,
\bar Q_3Q_4,\bar Q_1Q_4,$ and $Q_2\bar
Q_3$ terms at the Planck scale.
{}From the symmetry, we expect the 
nonrenormalizable term given in Eq.~(7),
which is the result of supergravity effects. 
But below the h-sector scale, even without the
nonrenormalizable term given in Eq.~(7), we may consider the effect of 
the  $S,S^\prime$ sfermion exchange diagram,
and the suppression factor is of order $M_P$ since $M$ in
Eq.~(7) is of order $M_P$ from the naturalness argument.
On the other hand, even without the nonrenormalizable term in Eq.~(7), 
$\partial W/\partial S=0$ gives $S^\prime=-\bar Q_1Q_2/M$ which,
after inserted in the $H_1H_2S^\prime$ term of (7), 
gives the desired nonrenormalizable
term below the h-sector scale.
In any case, below the h-sector scale we consider Eq.~(7).
With $<\bar Q_1Q_2>\sim \Lambda_h^2$, we obtain $\mu$ of
order the electroweak scale \cite{ckn}.

The singlet fields $S$ and $S^\prime$ are removed at the Planck
scale. At low energy there remain $H_1,H_2,\bar Q_1,
Q_2,\bar Q_3$, and $Q_4$. The h-gluinos can couple to
the h-quarks through the h-sector strong interactions
to give
$$
\int d^2\theta W^\alpha W_\alpha\left( {1\over 4}
+f(W^\alpha W_\alpha, {\rm\ Det\ }\bar QQ)\right)
$$
where the first factor comes from the h-gauge sector kinetic energy 
term and the second factor is the result of the h-sector dynamics and
is a function of two arguments respecting the 
global symmetry below the h-sector scale. This global symmetry
is $SU(N_f)_1\times SU(N_f)_2\times U(1)_B\times 
U(1)_C\times U(1)_{\tilde R}$ where 
$U(1)_C$ is anomalous and $U(1)_{\tilde R}$ is anomaly free.
These quantum numbers of the h-sector fields 
are given in Table 3. $C$ and $\tilde R$ are linear combinations
of $A$ and $R$. But for the study of h-sector dynamics,
$C$ and $\tilde R$ are more convenient. 
The h-sector scale $\Lambda_h$ has the nontrivial
$C$ transformation property to match anomaly \cite{yank}. In the
table, the composite meson field $T=\bar QQ$ is also shown.

\vskip 0.5cm
\centerline{Table 3. The $SU(N_f)_1\times SU(N_f)_2$, 
$C$ and $\tilde R$ quantum numbers.}
\vskip 0.5cm
\begin{center}
\begin{tabular}{|c|c|c|c|c|c|}
\hline
&\ $SU(N_f)_1$\ &\ $SU(N_f)_2$\ &\  $B$\ &\ $C$\ &\ $\tilde R$\ \\
\hline
\ \ $W^\alpha$\ \  &1&1&0&0&1\\
\ \ $Q_i$\ \  &$N_f$&1&1&1&$-(N_c-N_f)/N_f$\\
\ \ $\bar Q_i$\ \  &1&$\bar N_f$&\ \ --1\ \ &1&$-(N_c-N_f)/N_f$\\
\ \ $T$\ \  &$N_f$&$\bar N_f$&0&2&$-2(N_c-N_f)/N_f$\\
\ \ $\Lambda_h^{3N_c-N_f}$\ \  &1&1&0&$\ \ 2N_f\ \ $&0\\
\hline
\end{tabular}
\end{center}
\vskip 0.5cm
{}For $N_c=2$ and $N_f=1$, the 't Hooft instanton interaction
with $2N_c$ gluino lines and $2N_f$ quark lines is derived
from 
$$
S_{ins}= \int d^2\theta (W^\alpha W_\alpha)^2\left(
{{\rm Det}\bar QQ\over \Lambda_h^5}\right).
$$
Consistently with the global symmetry of Table 3, we can write
many terms of the form
\begin{equation}
\int d^2\theta 
\ (W^\alpha W_\alpha)^n\left({\Lambda_h^{3N_c-N_f}\over {\rm Det}\bar QQ}
\right)^{n-1\over N_f-N_c},
\end{equation}
where $n$ is a nonnegative integer.
However, the information on the anomaly matching fixes the relative
coefficient to give \cite{yank}
\begin{equation}
\int d^2\theta S\left[\log\frac{S^{N_c-N_f}{\rm Det}T}{c^{1/3}\Lambda_h^{3
N_c-N_f}}-(N_c-N_f)\right],
\end{equation}
where $S=W_\alpha W^\alpha$ and $T_{ij}=\bar Q_iQ_j$, and $c$ is
a number of order 1.  The symmetry dictates the form of 
the effective interaction to the above form, Eq.~(8), which
coincides with the 't Hooft interaction for $N_c=2, N_f=1$ and $n=2$.
Even though these cannot be generated perturbatively, but
can be generated nonperturbatively since these terms respect the
global symmetries \cite{witten,seiberg}. The relevant scale for
the nonperturbative generation of the above terms is the h-sector
scale $\Lambda_h$.

{}For illustration we take $N_c=3$ and $N_f=2$
corresponding to two flavors of Table 2.  
Below the h-sector scale, let us represent
\begin{equation}
S\equiv W^\alpha W_\alpha = m^2Z,\ \ T_{ij}\equiv 
\bar Q_iQ_j=m^\prime \Phi_{ij}
\end{equation} 
where $Z$ and $\Phi_{ij}$ are the effective chiral superfields,
and $m$ and
$m^\prime$ are at the h-sector scale. Let us apply $SU(N_f)_1
\times SU(N_f)_2$ transformation so that the matrix $\Phi$ is
diagonal $\Phi_{ij}=\Phi_i\delta_{ij}$.  Then the relevant 
superpotential below the h-sector scale is given by
\begin{equation}
W_{eff}=-m^2Z+m^2Z\log
\frac{m^2m^{\prime 2}Z\Phi_1\Phi_2}{c^{1/3}\Lambda_h^7}.
\end{equation} 
Minimization of $W_{eff}$ gives
\begin{equation}
Z\Phi_1\Phi_2=\frac{c^{1/3}\Lambda_h^7}{m^2m^{\prime 2}},\ Z=0,
\ \Phi_1=\Phi_2=\infty.
\end{equation}
This runaway solution is a typical feature of the massless
supersymmetric QCD. Therefore, in the massless theory, we
may have only a cosmological interpretation for a nonzero $Z$.

If two quarks obtain masses of $m_1$ and $m_2$, we add the following
terms in the effective superpotential, Eq. (11),
\begin{equation}
-m_1\bar Q_1Q_2-m_2\bar Q_3Q_4.
\end{equation} 
>From Eq. (7), we note $m_1\simeq <S>$ and $m_2=0$. After the
introduction of soft supersymmetry breaking terms at order $m_{3/2}$,
$S$ develops a vacuum expectation value of order $m_{3/2}\Lambda_h^2/M^2$,
which implies $m_1\sim 10^{-10}m_{3/2}$. 
The minimization of $W_{eff}$ gives 
\begin{equation}
\frac{m^2Z}{\Phi_1}-m_1m^\prime=0,\ \frac{m^2Z}{\Phi_2}-m_2m^\prime=0
\end{equation}
in addition to the first equality of Eq. (12). 
These cannot be satisfied simultaneously. Supersymmetry is broken.
Still $\Phi_2$ runs away. We will cutoff $\Phi_2$ at $M_P$, for which we
try to introduce a nonvanishing $m_2\sim 10^{-6}m_1$ by hand. For 
nonzero $m_2$, we can satisfy Eq. (14), and determine
\begin{equation}
Z\sim 10^{-11}m_{3/2}\sim\ 1\ {\rm eV},\ \Phi_1\sim 10^{-1}\frac{
\Lambda_h^2}{m^\prime}\sim 10^{12}\ {\rm GeV},\ 
\Phi_2\sim 10^5\frac{\Lambda_h^2}{m^\prime}.
\end{equation}
Since $\Phi_1\sim 10^{12}$~GeV, we obtain a ball park $\mu$.

Since both $U(1)_A$ and $U(1)_R$ symmetries are broken by
the vacuum expectation values of $\Phi_1, \Phi_2$ and $Z$, there result
two Goldstone bosons. One is the familiar invisible axion (or the very
light axion) \cite{axion} and the other is an R-axion. The R-axion
is the pseudo-Goldstone boson resulting from the 
breaking of the $U(1)_R$ symmetry. The model given in Table~2 
have the nonvanishing divergences for both the $J^A_\mu$ and $J^R_\mu$ 
currents,
\begin{eqnarray}
&\partial^\mu J^A_\mu= -\frac{4}{32\pi^2}F^\prime\tilde F^\prime\\
&\partial^\mu J^R_\mu= \frac{2N_c}{32\pi^2}F^\prime\tilde F^\prime
\end{eqnarray} 
where $F^\prime\tilde F^\prime$ is the h-sector gluon anomaly,
$(1/2)\epsilon^{\mu\nu\rho\sigma}F^{\prime a}_{\mu\nu}
F^{\prime a}_{\rho\sigma}$. Even if the h-sector scale is $\Lambda_h$, 
the instanton potential is multiplied by
a factor $m_{\tilde G}^{N_c}m_1m_2/
\Lambda_h^{N_c-1}$ where $m_{\tilde G}$, $m_1$, and $m_2$ are
the masses of the h-gluino, fermionic partners $\bar Q_1$
(and its partner $Q_2$) and $\bar Q_3$ (and its partner $Q_4$), 
respectively. $m_{\tilde G}$ is expected to be of the
electroweak scale order. 
A nonzero $m_2$ can occur from
the nonrenormalizable terms in $W$, but these are too small
to give Eq.~(14).
Possible terms in the K{\a}hler potential are more suppressed.
In Eq.~(13), we added $m_2\sim 10^{-6}m_1$ which is small enough
not to invalidate our symmetry argument.

Because $H_1$ and $H_2$ carry Peccei-Quinn quantum numbers,
Eq.~(16) has a QCD gluon anomaly if QCD quarks are included.
There are two decay constants, $<\Phi_1>$ and $<\Phi_2>$.
Because the h-sector instanton potential is very shallow,
the decay constant corresponding to the QCD potential is
the smaller one, $<\Phi_1>$ \cite{quint}. 
The resulting axion is the very light one \cite{axion}
with the decay constant $\sim\Phi_1$~\cite{quint}, and
its nature is of composite type \cite{comp} since the Peccei-Quinn
symmetry is broken by the h-squark condensation.\footnote{
The present model is much simpler than the intricate composite
model of Ref.~\cite{sul}.} 
The other pseudo-Goldstone boson is extremely light with decay
constant $\sim <\Phi_2>$.

\section{Conclusion}

In conclusion, we have emphasized that the solution of the
$\mu$ problem should have a root with the symmetry principle.
Along this line, the $U(1)_A\times U(1)_R$ symmetry is 
used to generate an electroweak scale $\mu$ naturally
from the dynamics of the hidden sector. This class of
models has a potential to house the much needed quintessence
since the symmetry we require may forbid the h-quark masses
down to a needed level \cite{quint}.
However, the example we presented here requires the
introduction of a nonzero parameter $m_2$. 
A more satisfactory solution would be to determine it dynamically.

\acknowledgments
I have benefitted from discussions with K. Choi, E. J. Chun,
H. B. Kim, and H. P. Nilles.
This work is also supported in part by KOSEF, MOE through
BSRI 98-2468, and Korea Research Foundation.

\end{document}